\documentstyle[11pt,newpasp,twoside]{article}
\def\edcomment#1{\iffalse\marginpar{\raggedright\sl#1\/}\else\relax\fi}
\reversemarginpar

\begin{document}
\title{Magnetic Fields are not ignorable in the dynamics of disks}
 \author{E. Battaner, E. Florido \& A. Guijarro}
\footnotesize
\affil{Dpto. F\'{\i}sica Te\'orica y del Cosmos, Universidad de Granada, E-18071 Granada, Spain}
\normalsize
\begin{abstract}
Magnetic fields are considered to be dominant when
$\varepsilon_{B}\geq\varepsilon_{K}$, being
$\varepsilon_{B}=B^{2}/8\pi$ the magnetic energy density and
$\varepsilon_{K}=1/2 \rho\theta^{2}$ the rotation energy density, for
a conventional moderate B= 1 $\mu$G. They are considered to be
negligible when $\varepsilon_{B}<\varepsilon_{K}$ for $B\sim 10
\mu$G. With no assumption and no theoretical calculation, we show that magnetic
fields cannot be ignored in the outer parts of a galaxy like the Milky
Way and in the whole disk of a dwarf galaxy.
\end{abstract}

\section{Description}
Magnetic fields are of the order of 5-10 $\mu$G in the interior of
a spiral galaxy (see, for instance, Beck 1991) and of the order of 1-3
$\mu$G in the intergalactic medium (see, for instance, Kronberg
1994). We therefore assume 1 $\mu$G in the outer region. This
interpolated value is therefore a conservative one.

\includegraphics{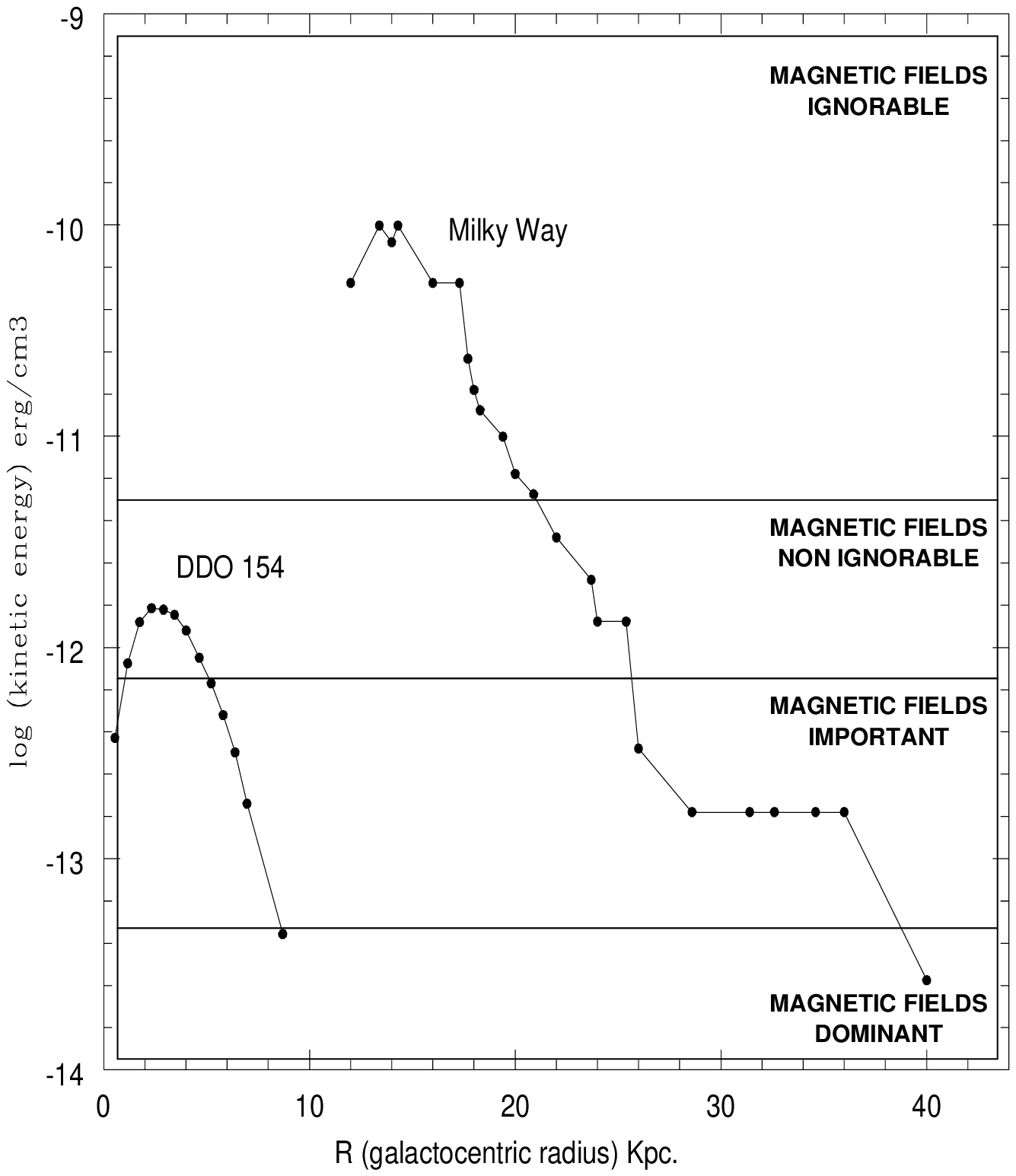}

\vspace*{9cm}
\noindent For the Milky Way, the density data have been taken
from  Diplas \& Savage (1991), (see also Burton 1992). For the
rotation curve it was assumed a constant value of 200 km/s (for R$\geq
R_{\odot}$), based on the works by Homma \& Sofue (1997), Merrifield
(1992), Olling \& Merrifield (2000).\\
For the typical irregular galaxy DDO154, the surface density and
the rotation data were taken from Gelato \& Sommer-Larsen
(1999). A typical thickness of 1 Kpc was assumed.\\           
A 1 $\mu$G in the outer disk does not produce detectable
synchrotron radiation, because \textbf{a)} under equipartition $I
\propto B^{7/2}$, i.e. the intensity, $I$, decreases much more faster
than B does. \textbf{b)} Some authors assume that the number density
of relativistic electrons is proportional to the density (because
their sources are supernova explosions, which are produced where
density is larger). In this case, $I$ decreases still faster, in an
exponential way, or even faster as a result of the typical stellar
truncation. Probably, B decreases much slower because the strength in the
intergalactic medium is $\sim 1\mu$G. \textbf{c)} The
synchrotron spectrum suddenly steepens for large radii (Lisenfeld et
al. 1996) what is easily interpreted as a truncation of relativistic
electron sources. 

\section{Conclusions}
When considering the rotation curve of spirals, magnetic
fields are not at all negligible. They are particularly important in
dwarf irregular galaxies where $\theta$ is lower. Magnetic
fields have similar values (see for instance, Beck 1991). Theoretical
models by Nelson (1988), Battaner et al. (1992), Battaner \& Florido (1995) and
Battaner, Lesch, \& Florido (1999), have already explored
this possibility.\\
\indent \textbf{Ignoring magnetic fields when interpreting rotation curves can
be completely unrealistic.}

\end{document}